\documentclass{article}

\usepackage{graphicx}
\usepackage{authblk}
\usepackage[caption=false]{subfig}
\usepackage{amsthm}
\usepackage{amsmath}
\usepackage{amssymb}
\usepackage[margin=1.25in]{geometry}
\usepackage{color}
\usepackage{morefloats}
\usepackage{epstopdf}

\theoremstyle{definition}
\newtheorem{mydef}{Definition}

\usepackage{hyperref}

\begin{document}

\title{Influence Activation Model: A New Perspective in Social Influence Analysis and Social Network Evolution}

\author[1,2]{Yang Yang}
\author[1,2]{Nitesh V. Chawla\thanks{Corresponding Author}}
\author[1,2]{Ryan N. Lichtenwalter}
\author[1,2]{Yuxiao Dong}
\affil[1]{Department of Computer Science and Engineering\\
University of Notre Dame\\
Notre Dame, IN 46556}
\affil[2]{Interdisciplinary Center for Network Science and Applications (iCeNSA), Department of Computer Science and Engineering\\
University of Notre Dame}

\maketitle

\begin{abstract}
What drives the propensity for the social network dynamics?  Social influence is believed to drive both off-line and on-line human behavior, however it has not been considered as a driver of social network evolution. Our analysis suggest that, while the network structure affects the spread of influence in social networks, the network is in turn shaped by social influence activity (i.e., the process of social influence wherein one person's attitudes and behaviors affect another's). To that end, we develop a novel model of network evolution where the dynamics of network follow the mechanism of influence propagation, which are not captured by the existing network evolution models. Our experiments confirm the predictions of our model and demonstrate the important role that social influence can play in the process of network evolution. As well exploring the reason of social network evolution, different genres of social influence have been spotted having different effects on the network dynamics. These findings and methods are essential to both our understanding of the mechanisms that drive network evolution and our knowledge of the role of social influence in shaping the network structure.
\end{abstract}

\section*{Introduction}
Social influence drives both offline and online human behavior and plays an especially important role in social sciences, for example, herding behavior in economics \cite{avery:1998}, rating behavior in financial market \cite{influence9}, and product recommendation in cultural markets \cite{salganik:2006}. In the domain of social influence analysis, the attitudes and tastes of individuals are believed to be influenced by others. Early models of social influence propagation were inspired by studies of epidemics, assuming that a piece of information could pass from one individual to another through social ties \cite{influence1, influence2, influence3, influence4, influence5, influence6, influence7, influence8}. Given the fact that the network structure affects the spread of influence among humans \cite{centola:2010}, it is still unknown that whether the social influence activity impacts the process of social network evolution. Here, we model how social influence can impact the shaping of social interactions and provide evidence that social influence is an important dimension underlying social network dynamics and resulting evolution. Our analysis provides a deeper understanding of the social network evolution, and has important implications for the study of the connection between social influence and social network evolution than previously considered.

In most existing work, the notion of social influence presupposes that individuals embedded in social networks are influenced by each other through social interactions. For example, John buys a product and his friend Doe is influenced by him and buys the same product. Such kind of adoption may cascade through the social ties in social networks, which is called \textit{locality influence} \cite{influence10, influence12, romero:11}. However, another genre of social influence called \textit{popularity influence} (i.e., global influence or external influence \cite{influence10, influence11}) still has considerable effects at different scales. The spread of \textit{popularity influence} does not necessarily rely on the network structure. For example, a famous person may have a considerable influence on individuals, no matter whether they have direct social ties or not. We define (i) \textit{locality influence} as the information diffusion or innovation adoption propagating through social ties, and (ii) \textit{popularity influence} as the influence that does not rely on social network topology and has global impact on individuals. The {popularity influence} has indispensable effects on the adoption of innovations as well as the spread of social behaviors \cite{influence10}. Our model incorporates both of {locality influence} and {popularity influence}, and considers them operating at different scales in the process of social network evolution.

Disentangling the mechanisms underlying the social network evolution is one of social science's unsolved puzzles. Much of existing models were devoted to reproducing the growth and evolution of network topology, and traditionally focused on defining basic principles driving link creation \cite{popularity_similarity, triad1,triad5,triad6,scalefree}. The connection between social influence and social network evolution is not yet explored. We posit that social influence capital (locality and popularity) impacts the growth of social networks as an intricate mechanism between locality influence and popularity influence, which may not be captured by the existing network evolution principles. Although previous research has not yet identified social influence as an origin of social network dynamics, there exists similarity between social influence {spreading} mechanisms and some existing network evolution principles. The notion of {locality influence} presupposes that individuals embedded in social networks are likely to be influenced by their friends. Cascade phenomena are well observed in the process of locality influence propagation, where individuals adopt a new action or idea locally \cite{zhang:13}. While Leskovec et al. \cite{triad1} identified the locality property of link creation, which echoes the locality influence mentioned above. The notion of popularity influence presupposes that popular individuals have larger impact over the social systems \cite{influence10, influence12}. Anecdotal evidence that preferential attachment \cite{scalefree} is a powerful mechanism underlying the emergence of scaling in social networks, where new links are established preferentially to more popular nodes, is ubiquitous \cite{popularity_similarity, triad5, triad6}. Although these similarities are subtle, they still provide clearly reasons that lead to our investigation into the relation between social influence and social network evolution.

\subsection*{Locality Influence}
\begin{figure}[ht]
	\centerline{\includegraphics[height=2.5in]{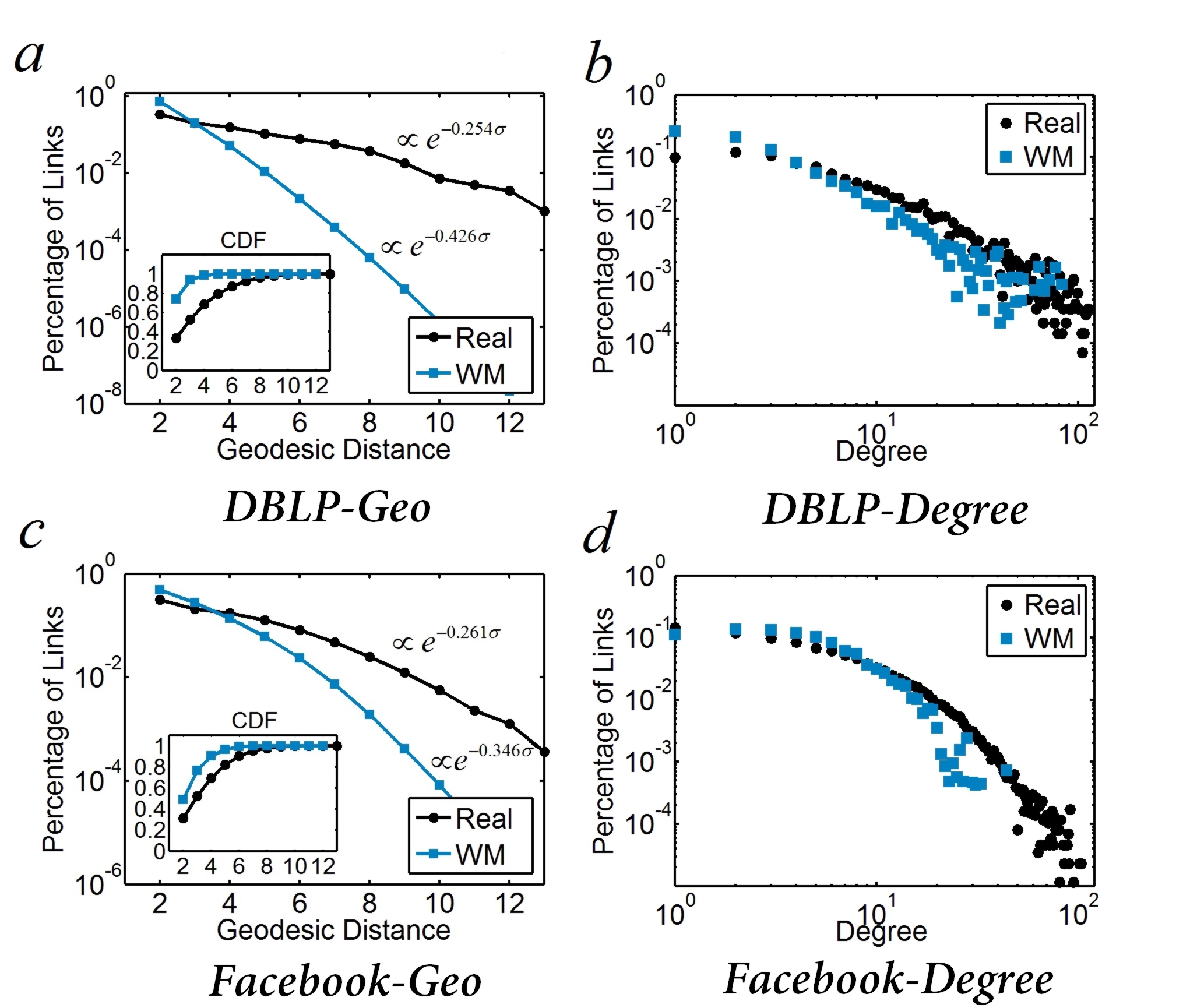}}
	\caption{The Link Formation Distribution Over Geodesic and Degree. In figures (a) (c) we compare the distributions of formed links over geodesic distance suggested by \textbf{WM} with the corresponding real link dynamics; in figures (b) (d) we provide the link formation distributions over degree for both \textbf{WM} with the real data.}
\label{fig_wm_simulation}
\end{figure}

Locality influence can have indispensable effect on the process of social dynamics. Due to the similarity between the spread of innovations and epidemic spreading, studies of social influence mainly employ epidemic branching processes to describe the cascade of locality influence on the social network \cite{influence1, influence3, influence4}. In order to better understand the effect of social influence on the evolution of social networks, we first hypothesize that the edge creation processes follow mechanisms of locality influence diffusion, and then validate whether the generated synthetic networks satisfy well with the real data. Among several existing influence diffusion models, we select \textit{weighted cascade model} for our validation \cite{influence1} {due to its free of parameters}. In our proposition, influence and link formation are closely interrelated in the process of network evolution. The link between two nodes forms due to the reason that their mutual influence is strong enough, while in turn the formation of link enhances the peer influence between these two nodes. We design a network evolution model, notated as \textbf{WM} (\textcolor{blue}{Supplementary Information 2}), derived from an existing influence spreading model {Weighted Cascade Model} \cite{influence1}. Figure~\ref{fig_wm_simulation} charts the evolution driven by \textbf{WM} vis-a-vis real network dynamics. These experiments are conducted on two real-world social networks: DBLP and Facebook (see \textcolor{blue}{Supplementary Information 1}). For DBLP dataset, we use the network until year 1992 as the base network (where geodesic distances and degrees are computed) and calculate the distributions of novel links formed between year 1993 and year 2006 over geodesic distance and degree; while for Facebook dataset, the base network is constructed using data until month 25 and we plot the distributions of novel links formed afterwards (from month 26 to month 52) over geodesic distance and degree.

\begin{figure}[t]
	\centerline{\includegraphics[height=1.3in]{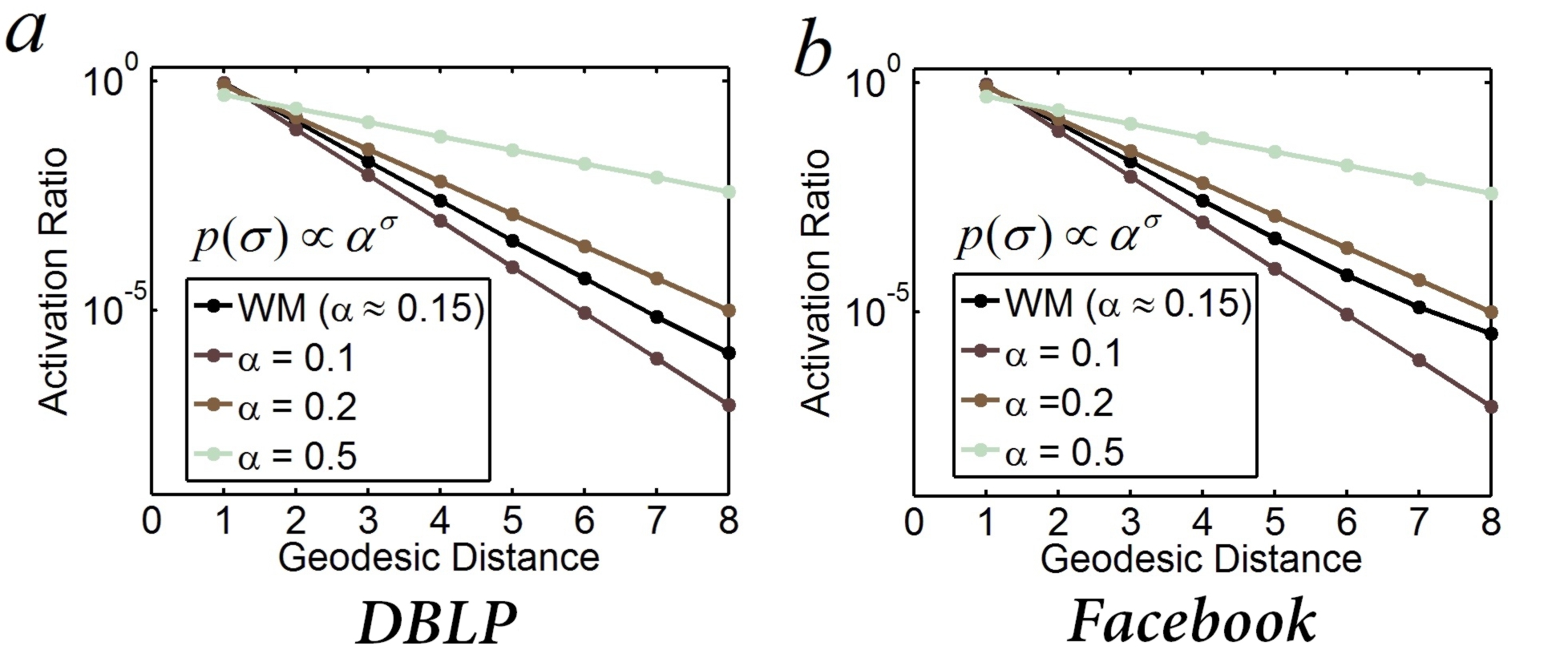}}
	\caption{Tuning $\alpha$. In this figure besides the curves of {\bf WM} model, we also provide the curves in the same family of distribution but with different parameters. Trivially we can observe that 1) the locality effect of {\bf WM} model follows exponential distribution, $p(\sigma) \propto e^{-log(\frac{1}{\alpha})\sigma}$; 2) the locality effect can be tuned by the parameter $\alpha$. The effect of locality increases when the parameter $\alpha$ decreases.}
\label{fig_alpha_tuning}
\end{figure}

A natural consequence of the locality effect is that the link formation probability decreases with the geodesic distance \cite{triad1, triad5, triad6, davidsen:2002}. The exponential decay suggests that the creation of a large fraction of edges can be attributed to the locality effect. Apparently, this phenomenon can possibly be explained by the social influence locality, which is validated in Figure~\ref{fig_wm_simulation} (a)(c). However the effect of {\it locality} is overestimated in the {\bf WM} model, which leads to its larger degree of locality compared to the real data. To quantitatively analyze the effect of {\it locality}, we take a closer look at the {\bf WM} model again. Based on the {\bf WM} model, we calculate the portion of nodes that can be successfully activated in each geodesic distance, which are given in Figure~\ref{fig_alpha_tuning}. In Figure~\ref{fig_alpha_tuning} we observe that it is best modeled as the exponential distribution:
\begin{align*}
p(\sigma) \propto \alpha^{\sigma} \rightarrow p(\sigma) \propto e^{-log({\frac{1}{\alpha}}) \sigma}
\end{align*}
The parameter $\alpha \simeq 0.15$ when the {\bf WM} model is applied in both DBLP and Facebook datasets. By knowing that the proportion of effective nodes ({nodes successfully activated by the source node}) in each geodesic distance follows exponential distribution ($p(\sigma) \propto \alpha^{\sigma}$), we can adjust the degree of locality by tuning the value of $\alpha$. The locality influence is more likely to activate close nodes (short geodesic distance). Additionally the degree of influence is able to be tuned quantitatively. This provides important implications in our further model design.

In Figure~\ref{fig_wm_simulation} (b) and (d) we observe that for the {\bf{WM}} derived distribution, high degree nodes ($d > 10^{1}$) form significantly lesser links than those actually formed in the networks (at least one order of magnitude difference). Thus, {\bf{WM}} does not {precisely} chart the macroscopic properties of a network, as dictated by preferential attachment, and a proxy for popularity influence. This implies that popularity influence is not appropriately represented in {\bf{WM}}. Furthermore, in Figure~\ref{fig_wm_simulation} (a) and (c), for each geodesic distance, we calculate the proportion $P(\sigma)$ of pairs $(u,v)$ separated by geodesic distance $\sigma$ that develop links. The synthetic $P(\sigma)$ simulated by {\bf{WM}} model resembles the real data. Both the real data and the {\bf WM} data are best modeled by the exponential distribution $P(\sigma) \propto \lambda exp(-\lambda \sigma)$. We have two observations: (i) the fact that both the real data and the {\bf WM} data are best modeled by exponential distribution implies the connection between locality of link formation \cite{triad1} and influence diffusion locality \cite{zhang:13}, (ii) we notice that the exponential parameters $\lambda$ of the real data and synthetic data are significantly different (67.7\% difference in DBLP and 32.5\% difference in Facebook). This indicates that \textit{locality} is just one dimension of the network evolution mechanism, and it is overrepresented in the \textbf{WM} model. In the work of \cite{weng:2013} the influence diffusion is mentioned to have impact on the link creation, however the locality influence effect is the only concern. As well demonstrating the correlation between locality influence and social network evolution, here we also provide evidence that only considering locality influence is not yet sufficient.

\subsection*{Popularity Influence}
Social influence analysis focus on local and endogenous processes such as word-of-mouth transmission, however they usually neglect equally important exogenous effects such as mass advertising. In general both endogenous and exogenous effects are present in social systems \cite{influence11}. The exogenous \textit{popularity influence} does not rely on the network structure for spreading, which has been considered in the case of the popularity of YouTube videos in the work of \cite{influence12}. Based on our observations in the above section, the locality influence itself is not sufficient enough to explain the process of social network evolution. We posit that the popularity influence also has considerable effects on the social network dynamics. In the vein of social network analysis, the preferential attachment model \cite{scalefree} postulates that new connections are made preferentially to more popular nodes (high degree nodes). While the work of social influence analysis consider popular individuals or products have higher influence in social systems. It is reasonable to incorporate popularity influence into the investigation of the relationship between social influence and social network evolution.

\begin{figure}[t]
	\centerline{\includegraphics[height=1.3in]{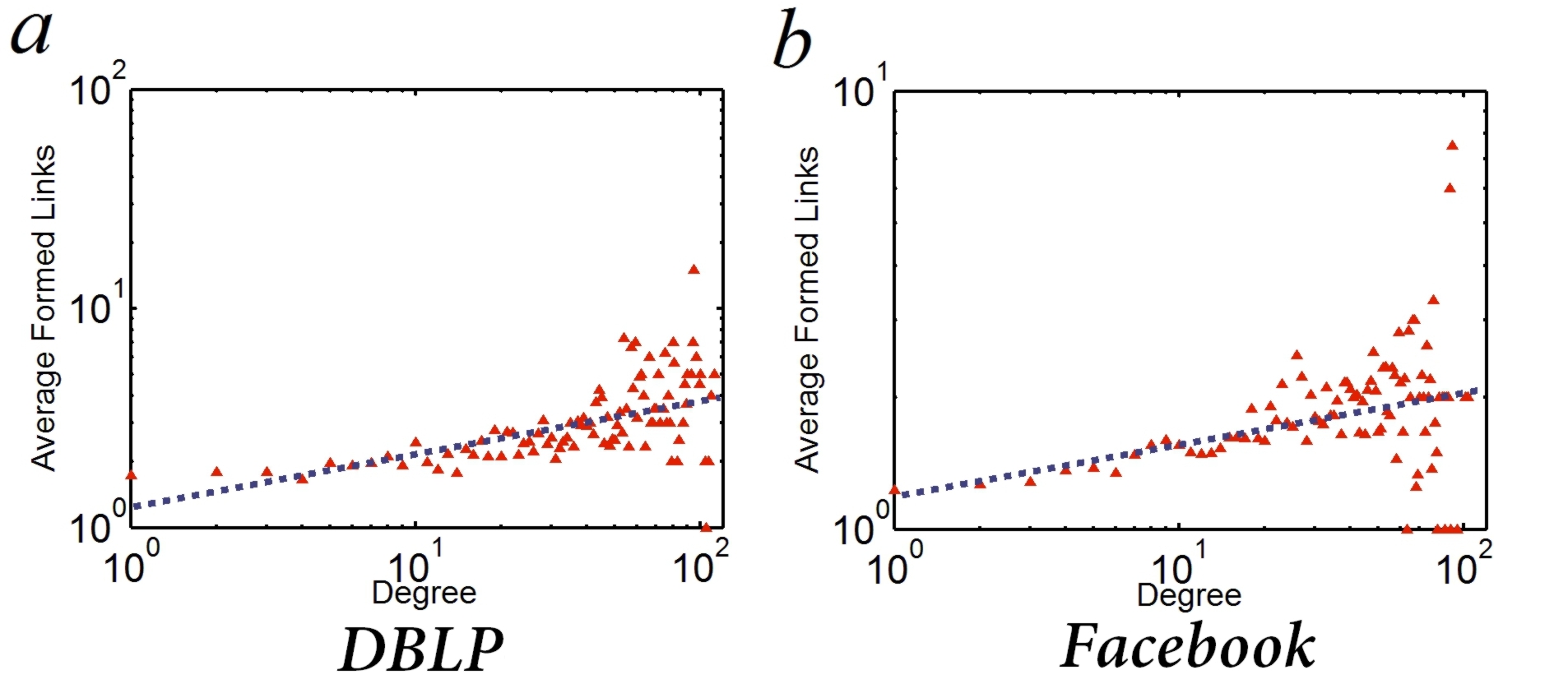}}
	\caption{Degree vs Average Link Formation Number. For nodes of degree $d$, we calculate the average number of novel links attached to them in the process of network evolution. This figure suggests that higher degree nodes are more likely to attracting new links.}
\label{fig_degree_link}
\end{figure}

\begin{figure}[ht]
	\centerline{\includegraphics[height=2.5in]{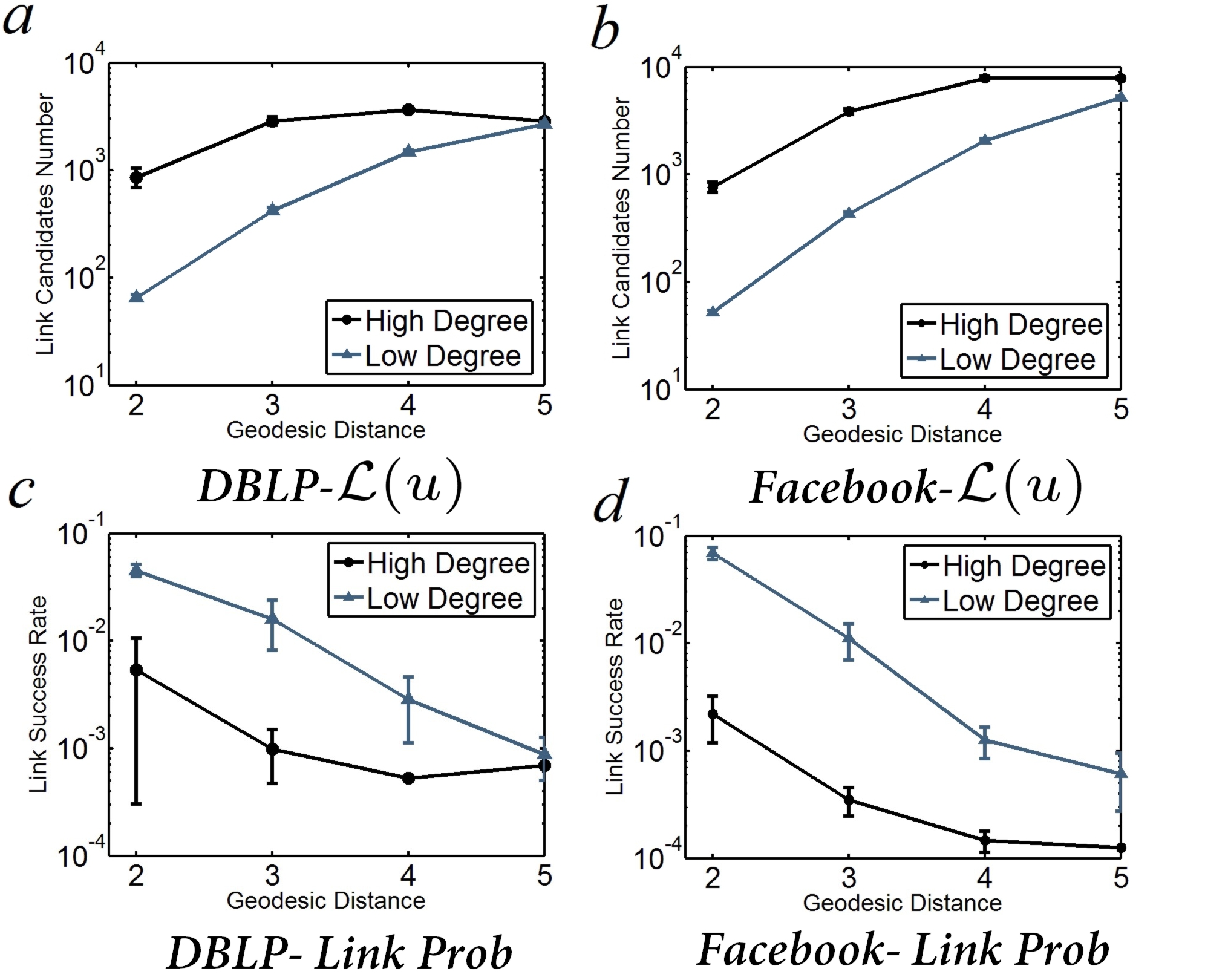}}
	\caption{Link Formation Differences between High Degree Nodes and Low Degree Nodes. a, b, We observe that the high degree nodes have larger number of {\it link candidates} than the low degree nodes. c, d, While the link success rate ($|\text{formed links}|/|\mathcal{L}(u)|$) of high degree nodes is much lower than that of low degree nodes. In DBLP dataset, we construct the network between year 1980 and year 1992 and identify the link candidates for high degree nodes and low degree nodes in each geodesic distance. By using the data between year 1993 and year 2006, we can compute the link success rate of high degree nodes and low degree nodes in each geodesic distance correspondingly. Similarly, for Facebook dataset, the link candidates are calculated based on the network between month 1 and month 25. The data between month 26 and month 52 are used to compute the link success rate.}
\label{fig_link_candidates}
\end{figure}

In Figure~\ref{fig_degree_link} we observe that the average number of links formed with high degree nodes is larger than that with the low degree nodes, which confirms the presence of {\it preferential attachment} \cite{scalefree} and implies that high degree nodes have a high volume of influence (if influence is driving the link creation). To further the understanding of this, we extract two subsets of nodes. Based on the ranking of node degree, we select the top 25\% quantile nodes and denote them as the set of {\it high degree nodes}; in the same way we pick the bottom 25\% quantile nodes as the set of {\it low degree nodes}. In Figure~\ref{fig_link_candidates} (a) and (b), we can see that the high degree nodes have larger number of {\it link candidates} (See \textcolor{blue}{Supplementary Information 3} and Figure~\ref{fig_link_candidates}) than the low degree nodes. Formally, the {\it link candidates set} is defined as follows:
\begin{mydef} {\bf Link Candidates Set}
For a node $u$ in network $G=\{V, E\}$, if $e(u, v) \notin E$ and there exists a path between $u$ and $v$, then the node pair $(u, v)$ is called a {\it{link candidate}} of node $u$ in network $G$. The set $\mathcal{L}(u) = \{ (u,v)| u,v \in V, e(u, v) \notin E \text{ and } \sigma(u,v) < \infty \}$ ($\sigma(u,v)$ is the distance between $u$ and $v$) is called the {\bf{link candidates set}} of node $u$.
\label{def_1}
\end{mydef}

However, the link success rate ($|\text{formed links}|/|\mathcal{L}(u)|$) of high degree nodes is much lower than that of low degree nodes (Figure~\ref{fig_link_candidates} (c) and (d)). These observations suggest that high degree nodes (i.e., implying popularity influence) have high volume of accumulative social influence but the unit influence is weaker. This also echoes with the finds of Dunbar in the work of \cite{dunbar:92}, where there is a limit to the number of people with whom one can maintain stable social relationships. This is important for our model design. We have two observations: first, high degree nodes have higher global influence throughout the whole network; second, the unit influence (influence on an individual candidate, $\phi (u,v)$) of high degree nodes is smaller than the low degree nodes, which leads to its low link success rate (\textcolor{blue}{Supplementary Information 3}). These findings will be incorporated into our model design.

\subsection*{Influence Activation Model}
Based on our observations and analysis above, we propose a novel model that incorporates the effects of popularity and locality, and is able to more accurately describe the evolution of social networks. Our model is a simulation of influence propagation and activation processes (combining popularity influence and locality influence). High influence activation probability from node $s$ to $t$ indicates high likelihood of link formation between them.

\begin{figure}[t]
\centerline{\includegraphics[height=2in]{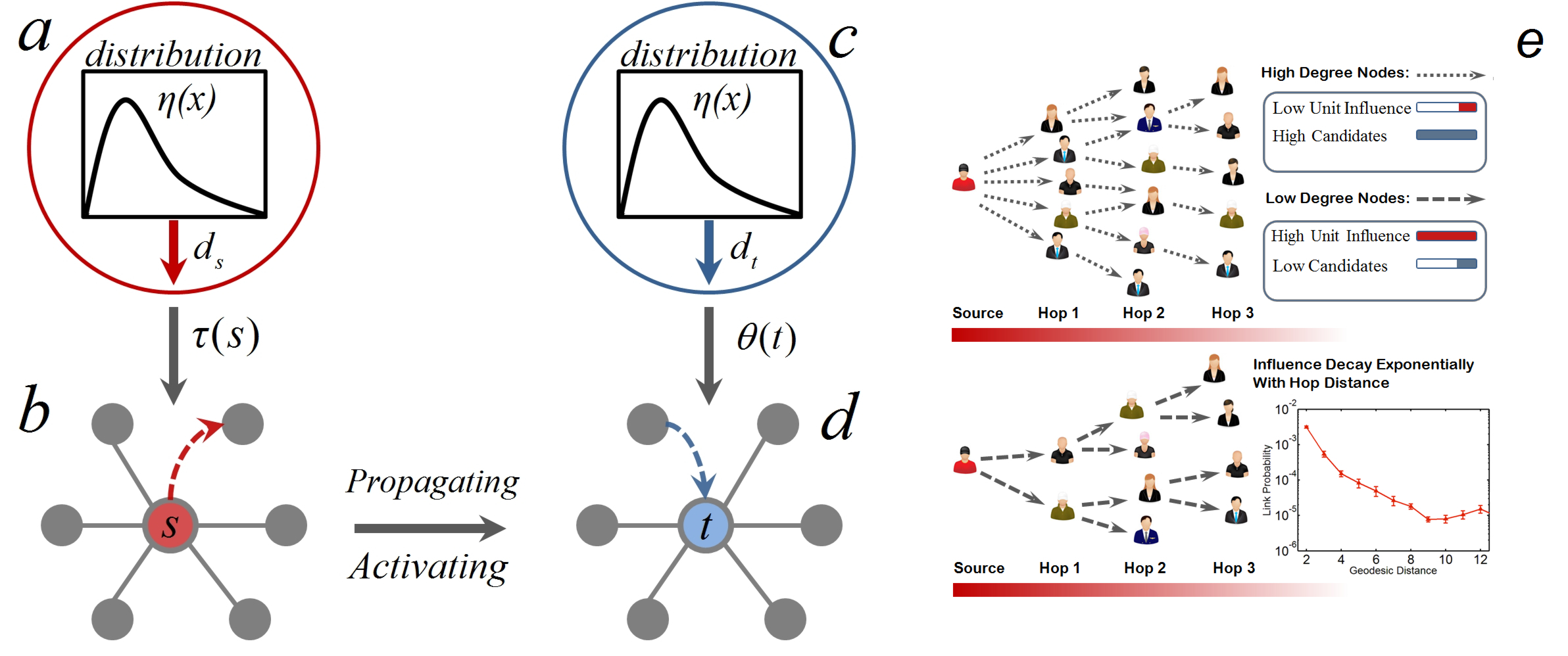}}
\caption{Propagation and Activation Process. In the heuristic of our IAM model (influence activation model), the source node $s$ is sending out flux of influence units, where each influence unit has an influence activation threshold $\tau(s)$. The value of $\tau (s)$ is the maximum number obtained after $d_{s}$ times random sampling from a distribution $\eta (x)$. In this way high degree nodes send out influence unit with high activation threshold, which
satisfies with our observation that high degree nodes have low success rate. At the same time, for the target node $t$ in the network, it has a corresponding value called {\it activating ability} $\theta (t)$. The condition that the target node is activated by the influence unit from source node is, $\tau (s) < \theta (t)$.}
\label{fig_activation_process}
\end{figure}

We postulate that for a node $s$, as the source node sending out an flux of \textit{identical} and \textit{independently} distributed influence units, an influence unit can only activate one node in its life time. Besides, in order to simulate the {\it popularity influence}, we define that the total volume of influence units sourced from node $s$ as proportional to its degree $d_{s}$ \cite{scalefree}. In our framework, the propagation and activation processes include two components: 1) the source node sends out the influence unit that propagates through the network; 2) the target node receives the influence unit and decides on being activated by the influence unit (Figure~\ref{fig_activation_process}). Based on our observations above, we define building blocks of our model accordingly (Figure~\ref{fig_activation_process}). For each influence unit originating from the source node $s$, we assume that it has an {\it activation threshold} $\tau (s)$. Thus high degree nodes have high {\it activation threshold} (weak unit influence power), and low degree nodes have small {\it activation threshold} (strong unit influence power) (Figure~\ref{fig_activation_process} (e)) (see \textcolor{blue}{Supplementary Information 3.3} and \cite{dunbar:92}). Similarly, we associate an {\it activating ability} $\theta (t)$ for the target node $t$. The node $t$ is successfully activated by the influence unit, when $\theta (t) > \tau (s)$ (See \textcolor{blue}{Supplementary Information 3}).

An influence unit originating from a node $s$ has a likelihood to activate any of the surrounding candidate nodes. Based on our observation the influence unit is more likely to activate nodes closer to it in terms of shorter geodesic distance (Figure~\ref{fig_activation_process} (e), locality). As we have observed in the above section, the portion of nodes that can be successfully activated in each geodesic distance can be best modeled as the exponential distribution. Thus we define the number of nodes that can be effectively activated by the influence unit from source $s$ within $\sigma$ hops as follows:
\begin{align}
r(s,\sigma) = \sum_{i=1}^{\sigma} |V_{i}| \times \alpha^{i-1}
\label{eqn_1}
\end{align}
where $V_{i}$ is the set of nodes which have $i$ hops geodesic distance from node $s$. The value of $r(s,\sigma)$ increases when the hop distance $\sigma$ increases. And the parameter $\alpha$ is used to tune the degree of locality effects (See \textcolor{blue}{Supplementary Information 3.3}). To note that, the parameter $\alpha$ is not set arbitrarily, it can be learned from the historical information of network dynamics (see \textcolor{blue}{Supplementary Information 4}). $r(s,\sigma)$ describes the effective number of nodes within $\sigma$ hops that are able to receive the influence signal sourced from $s$.

Based on the definition of propagation and activation process, we use the number of successful activations between $s$ and $t$ to estimate the volume of influence between them (\textcolor{blue}{Supplementary Information 3.3} for derivation):
\begin{align}
\phi(s,t)\propto \frac{d_{s}^{2}d_{t}}{(d_{s} + r(s,\sigma-1))(d_{s} + d_{t} + r(s,\sigma-1))}
\end{align}
As we presuppose that the social influence is highly correlation with the process of link creation, and we employ the volume of influence to estimate the link probability between $s$ and $t$:
\begin{align}
prob(s,t) \propto \phi(s,t)
\label{eqn_3}
\end{align}

We refer to it as influence activation model (IAM) (See \textcolor{blue}{Supplementary Information 3.3}), where the effects of locality and popularity are reconciled. $d_{s}$ and $d_{t}$ represent the degrees of nodes $s$ and $t$. $r(s,\sigma-1)$ describes the effective number of nodes within $\sigma-1$ hops that are able to receive the influence signal sourced from $s$, $\sigma$ is the geodesic distance between $s$ and $t$. The value of $r(s, \sigma-1)$ increases when the geodesic distance $\sigma$ increases. Obviously, the probability of link formation between $s$ and $t$ increases when the degrees of $s$ and $t$ increase, while decreases when the hop distance $\sigma$ increases. An implicit parameter in this model is $\alpha$ mentioned in Equation~\ref{eqn_1}, which can be learned from historical information of network dynamics (see \textcolor{blue}{Supplementary Information 4}). 

\section*{Results}
We consider two aspects for comprehensive evaluation --- macroscopic and microscopic validations. At the macroscopic level we consider various network characteristics that define the system or global level properties of a network. Macroscopic study of network focuses on network properties such as degree distributions, diameter, clustering coefficient, geodesic distribution, etc. At the microscopic level, we consider the aspect of link formation in a node's neighborhood, providing a perspective on the nature of human social interactions at a smaller scale  to understand the establishment and development of social relationships at a micro-level. Our goal is to verify the precision of the {\bf{IAM}} simulated network in imitating the growth observed in the real network, and compare to the established benchmark methods.

To stage the evaluation framework, we consider time-varying network in our evaluation (See \textcolor{blue}{Materials and Methods}). We compare {\bf IAM} to the Preferential Attachment model \cite{scalefree} ({\bf PA}) and another model {\bf WM} described above (See \textcolor{blue}{Supplementary Information 4.2}). There are two reasons to compare {\bf IAM} to the preferential attachment model and {\bf WM} model: (i) validate whether our model {\bf IAM} yields better description of network evolution; and (ii) verify whether the effects of popularity influence and locality influence are well reconciled in our model.

\subsection*{Macroscopic Validation} To illustrate that social influence is a strong force shaping the network structure and dynamics, and that our framework can precisely reconcile the popularity influence and locality influence, we compare the {\bf{IAM}} generated synthetic networks  with the real data in the macroscopic level. We consider a number of metrics that capture the macroscopic characteristics of the network. Our goal is to compare the characteristics derived from the synthetic network, generated by the respective models, against the actual network.

\begin{figure*}[t]
	\centerline{\includegraphics[height=2.5in]{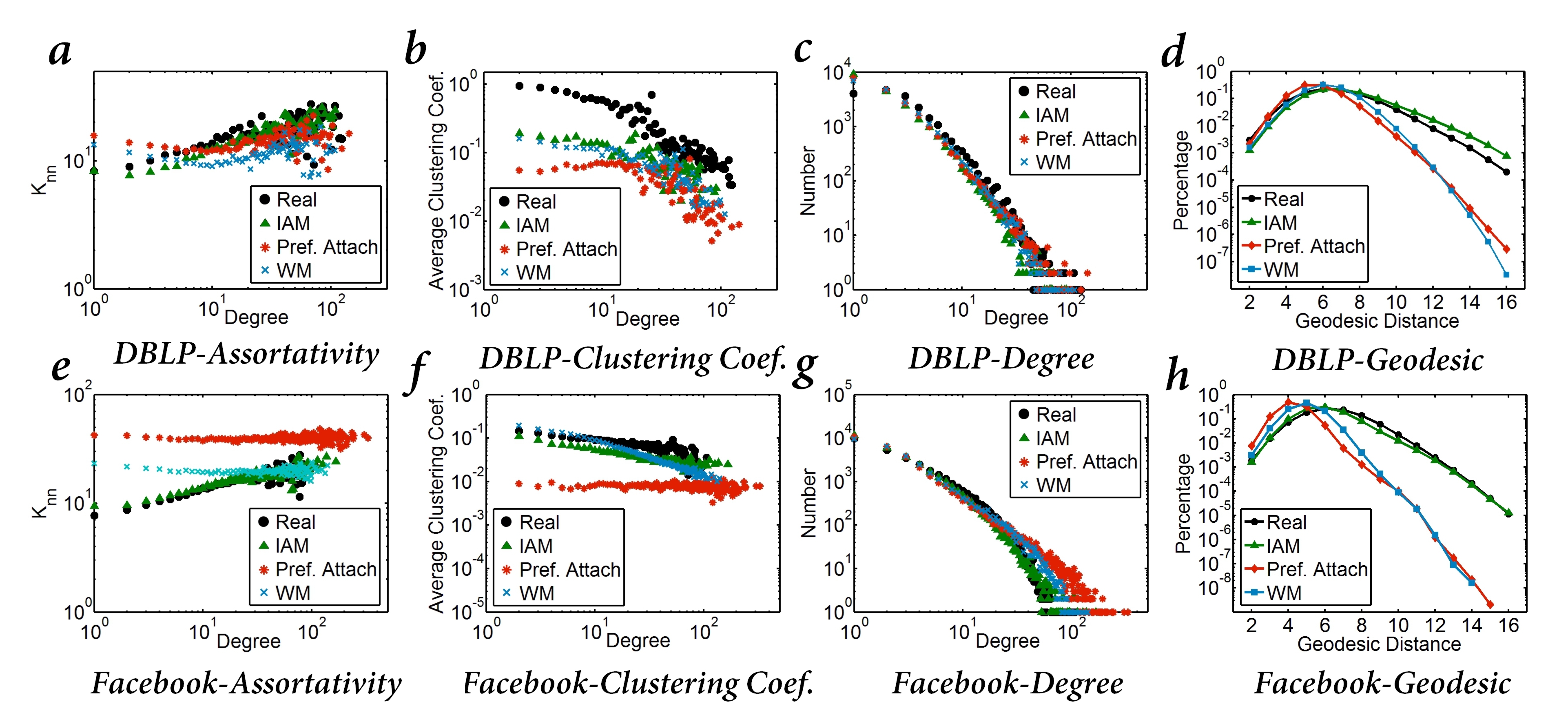}}		
	\caption{Macroscopic Properties Validation. a, e, $k_{nn}$ distribution, the average degree of neighbors of a node as a function of its own degree. This can depict the overall assortativity trend for a network. b, f, Clustering coefficient, we plot the average clustering coefficient as a function of degree, which for a node of degree $k$ measures the percentage of possible links between their $k$ neighbors (at most $k(k − 1)/2$) are present in their neighborhood graph. c, g, Degree distribution, the distribution of node degrees over the whole network. The degree distribution is very important in studying social networks. d, h, Geodesic distribution, the geodesic distribution of a network is the distribution of frequencies of geodesic distances - that is, the distribution of counts of the number of ordered pairs of nodes having each possible geodesic distance.}
\label{fig_evolution_performance}
\end{figure*}

In Figure~\ref{fig_evolution_performance} we provide the {\it $k_{nn}$ distribution} \cite{assortativity}, {\it clustering coefficient} \cite{clustering_coefficient}, {\it degree distribution} and {\it geodesic distribution} of the real data and three frameworks. In Figure~\ref{fig_evolution_performance} (a) (e) we observe that {\bf IAM} model successfully satisfies the property of assortativity, however the {\bf WM} model and the {\bf PA} model are not assortatitive. This implies that combining \textit{popularity influence} and \textit{locality influence} can yield a precise simulation of assortativity. In \cite{assortativity1} it has been validated that the assortative networks are more prone to the influence diffusion, and we demonstrate that the network structures generated by the influence driven model ({\bf IAM}) are also assortative. This further provides evidence that the social influence is the driving force shaping the network evolution. Additionally in Figure~\ref{fig_evolution_performance} (b) (f) we observe that {\bf IAM} and {\bf WM} yields higher clustering coefficients while the {\bf PA} model has much lower clustering coefficients than the real data. The reason is that the {\bf PA} model fails to simulate the locality property of link formation and then most of new links in {\bf PA} are formed in high geodesic distances. This leads to the low clustering coefficients of {\bf PA} model. However the locality property is well described by {\bf IAM} and {\bf WM}, where large portions of new links are formed in low geodesic distances. To notice that, in DBLP the {\bf IAM} model still has lower clustering coefficient than the real data. This is because in DBLP dataset the new links are introduce in the unit of cliques, which makes the clustering coefficients of DBLP large. If we remove the links between researchers who collaborate with each other only once, the differences between the real data and {\bf IAM} model shrink significantly.

In Figure~\ref{fig_evolution_performance} (d) (h) we further validate whether our model successfully simulates the real network geodesic distribution. Obviously both {\bf PA} and {\bf WM} fail to predict the correct geodesic distributions. The high geodesic distances node pairs are underestimated by the {\bf PA} and {\bf WM} models. This implies that the ignorance of locality ({\bf PA}) or only considering locality ({\bf WM}) will distort the geodesic distributions of networks. This further confirms the correctness of the {\bf IAM} model in predicting networks evolution.

Besides the validations of four above distributions, we also check other metrics. We perform more strict comparisons between synthetic data and real data. Graphlets are small connected non-isomorphic subgraph of a graph $G$ on $n \geq 3$ nodes of $G$. Graphlet degree distribution is confirmed to be informative for distinguishing different families of networks \cite{graphlet} where traditional distributions fail (i.e., {\it degree distribution}, {\it clustering coefficient}, {\it $k_{nn}$ distribution}, and {\it geodesic distribution}). We further compare synthetic networks with the real networks in terms of three metrics, graphlet degree distribution agreement \cite{graphlet}, relative graphlet frequency distance, and average path length difference. The results are provided in Table~\ref{tab_network_alignment}. Obviously the {\bf IAM} model generates synthetic networks that are more similar to the realistic networks in terms of graphlet measurements.

{\bf IAM}  effectively captures and reconciles the effects of popularity influence and locality influence, as measured by the macroscopic metrics. It also yields better performance than methods where either popularity or locality is modeled. Additionally it validates our propositions that social influence is a strong force driving the evolution of networks.

\subsection*{Microscopic Validation} We further explore the predictability of our models in microscopic level. To that end, we employ {\bf IAM} for link prediction in DBLP and Facebook. We employ the unsupervised approach to inferring new links in the near future, which is also described in the work of \cite{linkprediction1} \cite{linkprediction2}. For DBLP dataset, we employ the network from year 1980 to year 1992 as the training network and the network between years 1993 and 2006 as the testing network; while for Facebook dataset, the network from month 1 to month 25 is considered as the training network and the network between months 26 and 52 is used as the testing network. For the unconnected nodes pair $u$ and $v$ in training network, each model ({\bf PA} model, {\bf WM} model and {\bf IAM} model) generates a score $score(u, v)$ indicating the likelihood of link formation between them in future. {{Please note that our purpose is to simply validate the correctness of our model rather than solving the link prediction problem or comparing with other link prediction methods.}}

%Both DBLP and Facebook datasets allow us to evaluate the notion of popularity and locality for {\bf{IAM}}. For example, In DBLP, popularity prevails over locality, and the popularity of authors propagates outside of the network. While in Facebook, locality is the main factor driving the link creation.

\begin{figure*}[ht]
	\centerline{\includegraphics[height=2.5in]{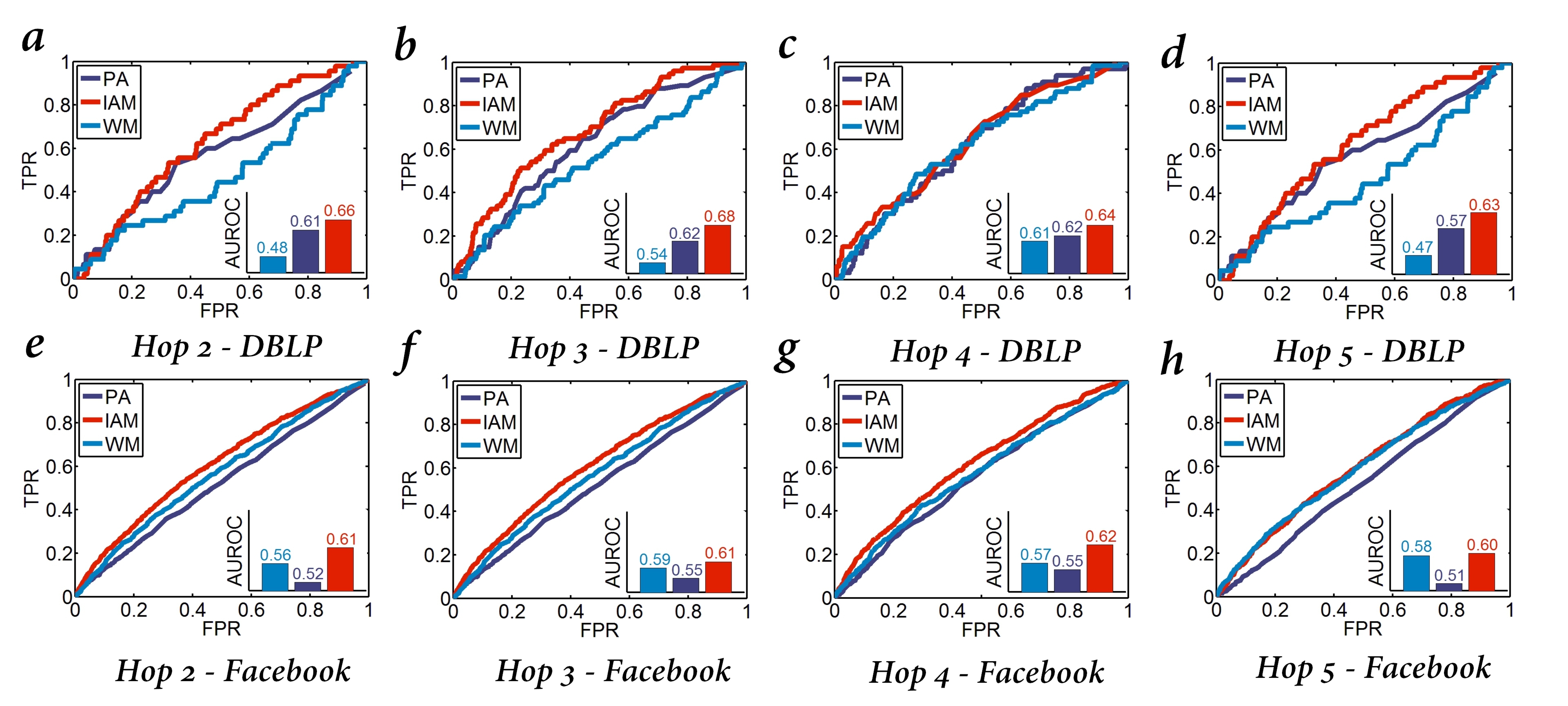}}		
	\caption{Inferring New Links. Top: The performance (ROC and AUROC) of three models in inferring new links over each geodesic distance (2-5) in DBLP; Bottom: The performance (ROC and AUROC) of three models in inferring new links over different geodesic distance (2-5) in Facebook}. 
\label{fig_lp_performance}
\end{figure*}

In Figure~\ref{fig_lp_performance} we provide the link prediction performance of {\bf IAM}, {\bf PA} and {\bf WM}. In order to provide more effective and fair evaluation of different link prediction methods \cite{linkprediction3}, we present the link prediction performance in different geodesic distances. And besides being measured in terms of AUROC, we also provide the AUPR performance (\textcolor{blue}{Materials and Methods}).

We have several interesting observations. First, we observe that in DBLP {\bf PA} generally has better performance than {\bf WM} and while in Facebook {\bf WM} yields better performance over {\bf PA}. This further confirms our proposition on differences in the evolution of social networks (from popularity and locality perspective). Secondly, in both DBLP and Facebook datasets our method {\bf IAM} has better performance than both {\bf PA} and {\bf WM} in terms of AUROC and AUPR. Again this identifies the ability of our {\bf IAM} model in describing the emergence of social networks from the microscopic perspective. It provides more evidence that {\bf IAM} effectively optimizes certain trade-offs between popularity and locality, which yields the best performance in circumstances where popularity and locality are in difference importance. In DBLP, popularity prevails over locality ({\bf PA} outperforms {\bf WM}). While in Facebook, locality is the main factor driving the link creation (Figure~\ref{fig_lp_performance}). To note that, in the work of \cite{christakis:09} Christakis et al. proposed that social influence functions mainly within 3 hops. And in our experiments we observe that IAM does not outperform a lot when the hop distance is large, to some extent this confirms the observations made in the work of \cite{christakis:09}.

\section*{Discussion}
Our work is the first to develop a unified model for network evolution, called the Influence Activation Model ({\bf{IAM}}), that captures the dynamics of network through the mechanism of influence propagation and link activation driven by social influence capital, which are not reflected by the existing network evolution models. We demonstrate that popularity influence and locality influence are key facets of social influence capital that govern network dynamics. Using different social networks and a variety of macroscopic and microscopic evaluation metrics, we show that {\bf{IAM}} is remarkably precise in charting the network evolution. That is, given a current time snapshot of a social network, we can effectively chart the path of network evolution not only from macroscopic network properties but also at the microscopic level to indicate neighborhood activity generated by link formation. The analytical framework behind {\bf{IAM}} and empirical evaluation demonstrate the importance of social influence in determining the scaling behavior and network topology of a time-varying network.

\section*{Methods}
\noindent{\bf Social Network Data}\\
Based on the DBLP dataset from \cite{dblp}, we choose authors who published at least 3 papers in conferences relating to four areas (Data Mining, Database, Information Retrieval, and Machine Learning) between 1980 and 2006. The {DBLP} dataset describes the collaboration relationship between academic researchers in computer science area. The DBLP network contains 26,136 authors and 87,191 collaboration relationships among them. The {Facebook} dataset is used by Viswanath et al.\cite{facebook}, which contains interactions among Facebook users between 2004-10 and 2009-01. There are 31,720 Facebook users and 654,424 links (friendship relationship) among them. Both of these datasets have the nuances of popularity influence and locality influence. In addition, both of these data sets have associated temporality so we can actually observe and validate the network evolution predicted by our Influence Activation Model.

\noindent{\bf Validation Methods}\\
To appropriately stage the evaluation framework, we consider time-varying network in our evaluation. For a time-varying network $G_{T}$, we are allowed to observe information from $G_{1}$ to $G_{\frac{T}{2}}$ and then evolve it from $t = \{ \frac{T}{2}, ......, T \}$ using a network evolution model to obtain a synthetic network $G'_{T}$. The objective of network evolution model is to maximize the similarity between the real network $G_{T}$ and the synthetic $G'_{T}$ \cite{triad1}. At the end of the evolution, we compare the macroscopic properties of $G_{T}$ and $G^{'}_{T}$. In DBLP datasets, we let the {\bf PA} model, {\bf WM} model and {\bf IAM} model evolve from the year 1993 to the year 2006; while in Facebook datasets, we let the simulation frameworks evolve from the month $26$ to the month $52$.

\noindent{\bf WM Model for Network Evolution}\\
In our proposition, influence and link formation are closely interrelated in the process of network evolution. In the \textbf{WM} model, we propose that the link likelihood between two nodes are proportional to the corresponding influence between them. In order to quantify the influence between two nodes in the network, we employ the methodology introduced in \textit{Weighted Cascade Model} \cite{influence1}. In \textbf{WM} model, the probability that node $v$ activates its neighbor $w$ (i.e., the chance that a person's attitudes and behaviors affect her/his neighbors) follows:
\begin{align*}
p_{v,w} = \frac{c_{v,w}}{d_{v}}
\end{align*}
where $d_{v}$ is the degree of node $v$ and $c_{v,w}$ is the weight of edge $e(v,w)$.

In order to measure the pairwise influence $\phi$ between two nodes, we employ breadth-first search procedure to propagate the probability of activating from the source node $u$ to any reachable node $w$ (see \textcolor{blue}{Supporting Information 2.1, Fig. S1}), the influence $\phi_{u,u}$ is initially assigned probability $1$. Throughout this procedure, the influence from the source $u$ on any reachable node will be recorded, and the influence $\phi_{u,w}$ can be computed as below:
\begin{align*}
\phi_{u,w} = max\{ \phi_{u,v}p_{v,w} \}, v \in \Gamma(w) \text{ and } \sigma_{u,v}+1 = \sigma_{u,w}
\label{eqn_wm}
\end{align*}
where $\Gamma(w)$ is the set of neighbors of node $w$ and $\sigma_{u,v}$ and $\sigma_{u,w}$ are the shortest path lengths from source node $u$ to $v$ and $w$ respectively.

After performing such a breadth-first search procedure for all nodes in the network, the influence between any pair of nodes can be computed. The networks analysed in this paper are undirected, thus the maximum value of $\phi_{u,w}$ and $\phi_{w,u}$ is considered as the mutual influence between nodes $u$ and $w$. In \textbf{WM} model, the probability of novel link between $u$ and $w$ is proportional to their mutual influence $\phi_{u,w}$:
\begin{align*}
prob(u,w) \propto \phi_{u,w}
\end{align*}

\noindent{\bf Network Measurements}\\
The measurements of networks are introduced as below:
\begin{itemize}
\item {\bf $k_{nn}$ distribution:} The average degree of neighbors of a node as a function of its own degree. This provides a way to capturing the overall assortativity trend for a network. If this function is increasing, the network is assortative, which shows that nodes of high degree connect, on average, to nodes of high degree; otherwise, the network is dissortative.

\item{\bf Clustering coefficient:} We calculate the average clustering coefficient as a function of degree, which for a node of degree $k$ measures the percentage of possible links between their $k$ neighbors (at most $k(k − 1)/2$) are present in their neighborhood graph. In most social networks the clustering coefficient decreases monotonically with degree, which is very important in social network analysis.

\item{\bf Degree distribution:} The degree of a node in a network is the number of edges connected to that node. The distribution of the fraction of nodes in the network that have each possible degree is called degree distribution.

\item{\bf Geodesic distribution:} The geodesic distance is the length of shortest path between two nodes. The geodesic distribution of a network is the distribution of frequencies of geodesic distances - that is, the distribution of counts of the number of ordered pairs of nodes having each possible geodesic distance.
\end{itemize}

\noindent{\bf Evaluation Metrics}\\
The performance of inferring new links are measured by ROC, Precision-recall Curve, AUROC and AUPR:
\begin{itemize}
\item {\bf ROC:} The receiver operating characteristic (ROC) represents the performance trade-off between true positives and false positives at different decision boundary thresholds.
\item {\bf AUROC:} Area under the ROC curve.
\item {\bf Precision-recall Curve:} Precision-recall curves are also threshold curves. Each point corresponds to a different score threshold with a different precision and recall value.
\item {\bf AUPR:} Area under the precision-recall curve.
\end{itemize}

\section*{Acknowledgments}
Research was sponsored by the Army Research Laboratory under Cooperative Agreement Number W911NF-09-2-0053, and by the grant FA9550-12-1-0405 from the U.S. Air Force Office of Scientific Research (AFOSR) and the Defense Advanced Research Projects Agency (DARPA)

\section*{Author Contributions Statement}
YY and NVC designed the research. YY and RL contributed analytic tools and performed empirical evaluation. YY, YD and NVC analyzed the results. YY and NVC wrote the paper. The authors declare no conflict of interest. NVC is the corresponding author: nchawla@nd.edu.

\section*{Additional Information}
Competing financial interests: The authors declare no competing financial interests.

\begin{table}[ht]
\caption{Network Alignments with Three Network Models. The highest correlation is in bold font. The results of graphlet degree distribution, RGF distance and path difference are generated by the GraphCrunch tool described in the work of \cite{graphlet} \cite{graphcrunch} and \cite{stickymodel}. The description of these metrics can be found in \textcolor{blue}{Supplementary Information 4}. In this table we can see in both datasets {\bf IAM} achieves better GDD-agreement than {\bf PA} and {\bf WM}. Additionally an interesting observation is, {\bf PA} has better GDD-agreement than {\bf WM} in DBLP while {\bf WM} has higher agreement value than {\bf PA} in Facebook. This further implies our proposition that locality influence in network and popularity influence have different effects on evolution of DBLP and Facebook. Additionally we observe that {\bf IAM} still achieves better performance than {\bf PA} and {\bf WM}.}
\begin{tabular}{@{\vrule height 10.5pt depth4pt  width0pt}clllllllll}
&\multicolumn3c{Graphlet Degree Distribution} & \multicolumn3c{RGF Distance} & \multicolumn3c{Path Difference} \\
\noalign{\vskip-11pt}
Networks\\
\cline{2-10}
\vrule depth 6pt width 0pt & IAM & PA & WM & IAM & PA & WM & IAM &PA& WM\\
\hline
DBLP & {\bf 0.900} & 0.897 & 0.888 & {\bf 1.623} & 2.800 & 2.169  & {\bf 0.845} & 1.009 & 0.891  \\\hline
Facebook & {\bf 0.953} & {0.907} & 0.935  & {\bf 1.792} & {2.269} & 1.810 & {\bf 0.417} & {2.126} & 1.491  \\\hline
\end{tabular}
\label{tab_network_alignment}
\end{table}

\end{document}